\documentclass[%
 reprint,
 amsmath,amssymb,
 aps,
 prb,
 floatfix,
 footnoteinthebib,
 longbibliography
]{revtex4-2}

\usepackage{graphicx,physics}% Include figure files
\usepackage{dcolumn}% Align table columns on decimal point
\usepackage{latexsym}
\usepackage[normalem]{ulem}
\usepackage{amssymb}
\usepackage{url}
\usepackage{soul}
\usepackage{verbatim}
\usepackage{multirow}
\usepackage{amsmath}
\newcommand{\beq}{\begin{eqnarray}}
\newcommand{\eeq}{\end{eqnarray}}
\usepackage{mathrsfs}
\usepackage{soul}
\usepackage[dvipsnames]{xcolor}
\usepackage{mathtools}
\usepackage{slashed}
\usepackage{epstopdf}
\usepackage{subfigure}  % use for side-by-side Fig.ures
\usepackage{hyperref}   % use for hypertext links, including those to external documents and URLs
\usepackage{bbold}
\usepackage{wasysym}
\usepackage{feynmp}
\usepackage{float}
\hypersetup{colorlinks,
}
\usepackage{bm}%varepsilon
\usepackage{pifont}
\usepackage[latin1]{inputenc}

\begin{document}

\title{\Large Topological Magnetic Phases and Magnon-Phonon Hybridization\\ in the Presence of Strong Dzyaloshinskii-Moriya Interaction}

\author{Weicen Dong$^{1,2,3}$}
\author{Haoxin Wang$^4$}
\author{Matteo Baggioli$^{1,2,3}$}
\email{b.matteo@sjtu.edu.cn}
\author{Yi Liu$^{5,6}$}
\email{yiliu42@shu.edu.cn}
\address{$^1$School of Physics and Astronomy, Shanghai Jiao Tong University, Shanghai 200240, China}
\address{$^2$Wilczek Quantum Center, School of Physics and Astronomy, Shanghai Jiao Tong University, Shanghai 200240, China}
\address{$^3$Shanghai Research Center for Quantum Sciences, Shanghai 201315,China}
\address{$^4$Department of Physics, The Chinese University of Hong Kong, Shatin, New Territories 999077, Hong Kong, China}
\address{$^5$Institute for Quantum Science and Technology, Shanghai University, Shanghai 200444, China}
\address{$^6$Department of Physics, Shanghai University, Shanghai 200444, China}

\begin{abstract}
In recent years, the interplay between quantum magnetism and topology has attracted growing interest, both for its fundamental importance and its technological potential. Topological magnons, quantized spin excitations with nontrivial band topology, hold particular promise for spintronics, offering routes to robust, low-dissipation devices for next-generation information processing and storage. While topological magnons in honeycomb ferromagnets with weak next-nearest-neighbor Dzyaloshinskii-Moriya interactions (DMI) have been extensively investigated, the strong-DMI regime remains largely unexplored. In this work, we examine topological magnetic phases and magnon-phonon hybridization in a two-dimensional magnetic system with strong DMI. We show that strong DMI drives a transition from a ferromagnetic ground state to a 120$^\circ$ noncollinear order. An additional Zeeman field further induces noncoplanar spin textures, giving rise to a diverse set of topological phases. We demonstrate that these topological phases can be directly probed through the anomalous thermal Hall effect. Finally, we find that the spin-spin interactions in the strong-$D$ phase enable magnon-phonon coupling that yields hybridized topological bands, whereas such coupling vanishes in the weak-$D$ phase.

\end{abstract}

\maketitle

\section{Introduction}
Magnons are quantized spin waves that can carry and process information without generating the Joule heat associated with electric currents~(e.g., \cite{Chumak2014,Mengqian2025}). Owing to this advantage, they have recently drawn considerable attention for their potential in next-generation electronic and computing technologies~\cite{Chumak2015}.

Meanwhile, topological materials \cite{Yan_2012}, such as topological insulators \cite{RevModPhys.82.3045}, have also been widely explored because of their robust physical properties, that emerge from their topological nature and can be possibly exploited for electronics \cite{gilbert2021topological} and other applications \cite{he2022topological}. In general, these topological excitations can be of different nature, including vibrational degrees of freedom (topological phonons) \cite{Tang2022,Yuanfeng2024}, optical ones (topological photons) \cite{PhysRevB.107.195104}, etc.

Magnetic materials can host nontrivial topological features~\cite{bernevig2022progress} and support topological magnonic excitations~\cite{annurevMcClarty} as well. Topological magnonic crystals exhibit protected chiral edge modes and hold strong potential for spintronic applications, such as spin-current splitters based on interfaces between distinct topological phases~\cite{PhysRevB.95.014435,PhysRevB.87.174427,PhysRevB.90.024412}.

In recent years, both experimental~\cite{zhang2024,PhysRevLett.123.080501,PhysRevB.106.L060408} and theoretical~\cite{PhysRevB.87.144101,PhysRevB.97.245111,PhysRevB.107.024408} efforts have greatly advanced the study of topological magnons. Theoretically, two-dimensional (2D) ferromagnetic honeycomb lattices with weak next-nearest-neighbor (NNN) Dzyaloshinskii-Moriya interactions (DMI) have been shown to realize magnonic Chern insulators hosting chiral edge states~\cite{Owerre_2016,Owerre20162,PhysRevLett.127.217202,Zhu2023,PANTALEON2018191,PhysRevB.106.104430,PhysRevX.11.021061,PhysRevMaterials.7.084402}. Experimentally, signatures of topological magnons with weak DMI have been reported in honeycomb ferromagnets such as CrBr$_3$~\cite{PhysRevB.104.L020402}, CrI$_3$~\cite{PhysRevX.8.041028}, CrSiTe$_3$, and CrGeTe$_3$~\cite{Fengfeng2021}.

Moreover, the DMI strength can be tuned through various means, such as inserting Ir layers~\cite{PhysRevB.90.020402}, applying electric fields~\cite{Srivastava2018}, modifying substrates~\cite{PhysRevB.106.L100407}, or introducing strain~\cite{PhysRevB.108.134405,PhysRevLett.127.117204}. Despite these advances, systematic studies of topological magnonic phases in the strong DMI regime remain largely unexplored.

In addition, the magnon-phonon coupling has been extensively investigated in recent years~\cite{bozhko2020magnon,PhysRevB.99.184442,PhysRevB.96.174416,PhysRevLett.123.167202,PhysRevLett.124.147204}. The combination of Heisenberg exchange and out-of-plane DMI (parallel to the spin direction), as discussed in Refs.~\cite{Owerre_2016,PhysRevB.104.L020402,PhysRevX.8.041028,Fengfeng2021}, induces magnon-phonon coupling that gives rise to higher-order bosonic interactions beyond linear spin-wave (LSW) theory. In contrast, in-plane DMI, oriented perpendicular to the spins, primarily alters the band topology~\cite{PhysRevB.104.045139,PhysRevB.106.214424,2023Baosong,PhysRevB.104.064305}. Since most prior studies have focused on collinear magnetic ground states, a natural and open question arises: how do strong DMI modify this picture?

In this work, we investigate a minimal model of a 2D magnetic system subject to both an external magnetic field (Zeeman term) and out-of-plane DMI, exploring the full range of DMI strength, from weak to strong coupling. By systematically varying the DMI coupling $D$ and the Zeeman field $h$, we construct the corresponding phase diagram and uncover a rich landscape of magnetic phases with distinct topological characteristics. We show that these topological phases, as well as the transitions between them, can be directly probed through the thermal Hall effect. Furthermore, we show that the distance-dependent spin-spin interaction mediates magnon-phonon hybridization, giving rise to topologically nontrivial hybrid bands. The quadratic magnon-phonon coupling appears only in the strong-$D$ phase and vanishes in the weak-$D$ phase.

\section{Model}

\begin{figure}
\includegraphics[width=\linewidth]{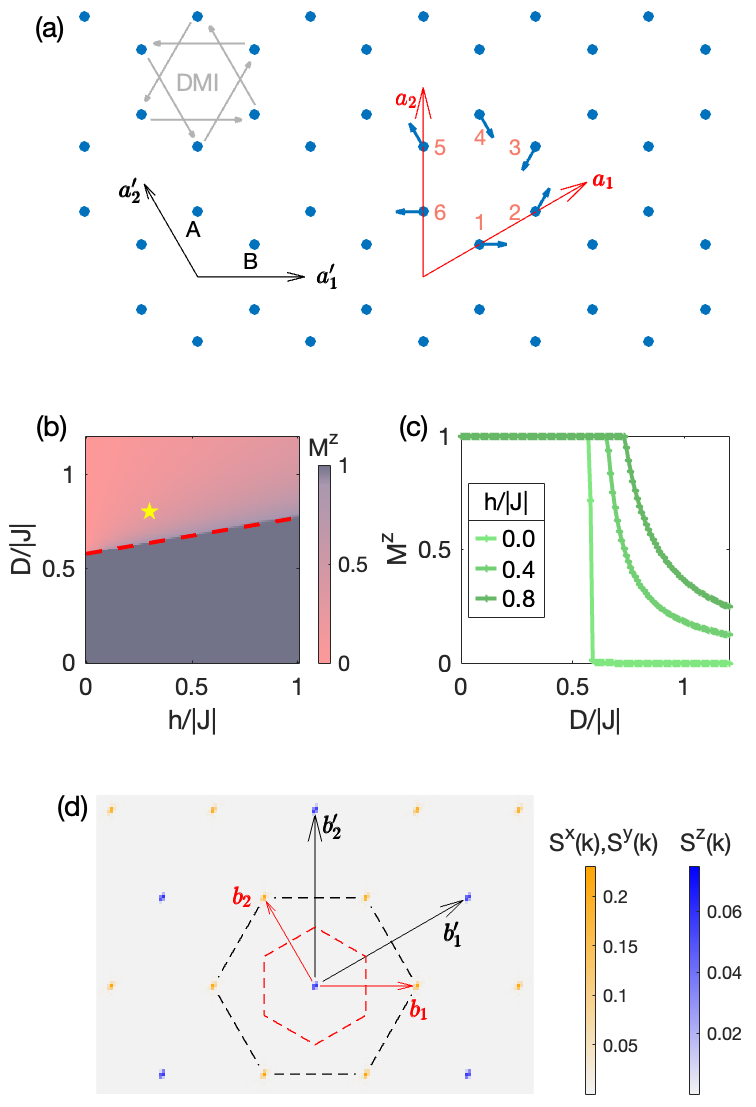}
\caption{\textbf{Classical ground state phases.} \textbf{(a)} Gray arrows represent the DMI. Black vectors $\boldsymbol{a}_1^\prime$ and $\boldsymbol{a}_2^\prime$ indicate the small unit cell for the weak-$D$ phase. Red vectors $\boldsymbol{a}_1$ and $\boldsymbol{a}_2$ mark the large unit cell for the strong-$D$ phase. Within the large unit cell, blue arrows show an example of 6-spin configuration in the strong-$D$ phase with $\phi_1=0$, $\phi_2=\frac{\pi}{3}$, and $D>0$.
\textbf{(b)} Ground-state order parameter $M^z$ as a function of $D$ and $h/|J|$. The red dashed line marks the analytic phase boundary, $D_c$ from Eq.~\eqref{criti}. \textbf{(c)} Order parameter $M^z$ versus $D$ for selected values of $h$. \textbf{(d)} In-plane components of the spin structure factor, $S^x(\textbf{k})=S^y(\textbf{k})$, and $z$ component, $S^z(\textbf{k})$, in the strong-$D$ phase. Black and red arrows indicate the unit vectors in $\boldsymbol{k}$ space, for the weak-$D$ and strong-$D$ phases, respectively. Black and red dashed lines indicate the Brillouin zone for the weak-$D$ and strong-$D$ phases, respectively. The parameters used are $D/|J|=0.8$ and $h/|J|=0.3$, corresponding to the yellow star in the phase diagram, panel (b).}
    \label{fig:1}
\end{figure}

We consider a 2D honeycomb lattice with nearest-neighbor (NN) ferromagnetic Heisenberg interactions, NNN DMI and Zeeman energy. The Hamiltonian is given by
\begin{equation}
\begin{aligned}
H=&J\sum_{\langle ij\rangle }\vec{S}_i\cdot\vec{S}_j+D\sum_{\langle\langle ij \rangle \rangle} (S_i^xS_j^y-S_i^yS_j^x)-h\sum_i S_i^z,
\end{aligned}\label{hami}
\end{equation}
where $\vec{S}_i$ is the spin on the $i$-th lattice site, $J<0$ the ferromagnetic coupling, $D$ the Dzyaloshinskii-Moriya coupling and $h$ the Zeeman coupling. Moreover, $\langle \dots\rangle$ indicates summation over NN indices, while $\langle\langle\dots\rangle\rangle$ over NNN ones. In the analytical derivations, the variables $S=|\vec{S}|$ and $J$ are kept explicit, while, without loss of generality, in the numerical calculations we set $S=1$ and $J=-1$ meV. Finally, we assume that LSW theory~\cite{Toth_2015,auerbach2012} is valid and higher-order corrections in the bosonic operators can be neglected. 

\subsection{Classical ground states}
Before exploring the whole phase diagram, we first analyze two useful limiting cases analytically. In the limit $D/J \to 0$, the ground state is ferromagnetic, with all spins align with the external magnetic field. The unit cell, containing two atoms, is denoted by $\boldsymbol{a}_1'$ and $\boldsymbol{a}_2'$, as illustrated in Fig.~\ref{fig:1}(a). On the other hand, in the opposite limit $D/J \to \infty$, the honeycomb lattice decouples into two independent triangular lattices composed of sites $(1,3,5)$ and $(2,4,6)$, respectively, as denoted in Fig.~\ref{fig:1}(a). We choose an enlarged magnetic unit cell with six atoms as a basis, defined by $\boldsymbol{a}_1$ and $\boldsymbol{a}_2$. In this limit, spins are expressed as $\vec{S}_i = S(\sin\theta\cos\phi_i,\, \sin\theta\sin\phi_i,\, \cos\theta),$
with the azimuthal and polar angles fixed by 
\begin{equation}
\begin{aligned}
\phi_3-\phi_1 &= \phi_5-\phi_3 = \phi_4-\phi_2 = \phi_6-\phi_4 
= \mathrm{sgn}(D)\,\tfrac{4\pi}{3}, \\
\cos\theta &= \frac{h}{3S(J+\sqrt{3}|D|)}.
\end{aligned}\label{eq:gs}
\end{equation}
This configuration corresponds to a $120^\circ$ noncollinear order in the $xy$ plane with a uniform canting angle $\theta$ relative to the magnetic field.
The spin orientations are illustrated by the blue arrows in Fig.~\ref{fig:1}(a). The in-plane Heisenberg interaction energy, $J\sum_{\langle ij \rangle }(S_i^xS_j^x+S_i^yS_j^y)$, vanishes for this configuration, resulting in degeneracy with respect to $(\phi_2-\phi_1)$. Meanwhile, $J\sum_{\langle ij \rangle}S_i^zS_j^z$ changes the value of $\cos\theta$, from $\frac{h}{3S\sqrt{3}|D|}$ to $\frac{h}{3S(J+\sqrt{3}|D|)}$.

Between the above two limits, we determine the classical ground state phases by numerically minimizing the classical energy. Starting from random spin configurations, we align each spin $\vec{S}_i$ along its effective magnetic field, $\vec{h}_{\mathrm{eff}}=-\partial H/\partial \vec{S}_i$. The out-of-plane average magnetization $M^z=(1/N)\sum_i S_i^z$ serves as the order parameter distinguishing different phases, where $N$ is the total number of sites. In Fig.~\ref{fig:1}(b), we present the calculated density map in the $(h,D)$ plane that allows us to draw a phase diagram for our magnetic system. We consider $D/|J| \in [0,1.2]$ and $h/|J| \in [0,1]$, where $h = g\mu_B B$ implies that $h/J = 1$ corresponds to $B \approx 8.6$ T. The boundary between the weak- and strong-$D$ phases, $D_c(h)$, corresponds to a first-order transition at $h=0$ and a smooth crossover at finite $h$ [see Fig.~\ref{fig:1}(c)].

For the weak-$D$ phase, $|D|\leqslant D_c(h)$, the dark gray region in Fig.~\ref{fig:1}(b), the system exhibits ferromagnetic order with $M^z=\mathrm{sgn}(h)$ (taken as $1$ for $h=0$). This state preserves the Hamiltonian $U(1)$ symmetry that corresponds to global spin rotations around the $z$ axis. 

For the strong-$D$ phase, $|D|>D_c(h)$, $M^z$ drops to zero when $D$ increases, indicating the breaking of the aforementioned U(1) symmetry and the transition to a new phase. The numerical result of spin configuration in strong-$D$ phase is exactly same as Eq.~\ref{eq:gs}. 

The critical line, $D_c(h)$, can be analytically derived by comparing the classical energies of the two phases. Within the enlarged unit cell, the classical energies are given by $E_{\text{weak}} = 9JS^2 - 6|h|S$ for the weak-$D$ phase, and $E_{\text{strong}} = 9JS^2 \cos^2\theta - 9\sqrt{3}|D|S^2 \sin^2\theta - 6hS\cos\theta$ for the strong-$D$ phase, respectively. The condition $E_{\text{weak}} = E_{\text{strong}}$ defines the critical line as
\begin{equation}
|D|=D_c(h) = \frac{-J}{\sqrt{3}} + \frac{|h|}{3S\sqrt{3}}. \label{criti}
\end{equation}
It is represented by the dashed red line in Fig.~\ref{fig:1}(b), demonstrating a good agreement between the analytical expression, Eq.~\eqref{criti}, and the numerical estimates.

In order to show the $120^\circ$ order in the strong-$D$ phase, choosing the parameters marked by a star in Fig.~\ref{fig:1}(b), we evaluate the spin structure factor, $S^\mu(\boldsymbol{k}) =  \Big| \frac{1}{N/2}\sum_i S_{i}^\mu e^{-i \boldsymbol{k} \cdot \boldsymbol{r}_i} \Big|^2$, where $\mu=x$, $y$, and $z$. The sum of $\boldsymbol{r}_i$ runs over the sites in the sublattice A or B, yielding equivalent results. As shown in Fig.~\ref{fig:1}(d), the out-of-plane component $S^z(\boldsymbol{k})$ (blue dots) exhibits peaks at all $\Gamma$ points of the Brillouin zone defined by $\boldsymbol{b}_1',\boldsymbol{b}_2'$, indicating a uniform spin alignment along the $z$ direction. In contrast, the in-plane components $S^{x}(\boldsymbol{k})=S^{y}(\boldsymbol{k})$ (orange) peak at $\boldsymbol{k}=k_1^\prime\boldsymbol{b}_1^\prime+k_2^\prime\boldsymbol{b}_2^\prime$ with $(k_1^\prime,k_2^\prime)=\pm(1/3,1/3)$. More specifically, the spins on sublattice A or B each form a single-$\boldsymbol{Q}$ state~\cite{Toth_2015}, where $\boldsymbol{Q}$ is denoted as $(k_1',k_2')=(Q,Q)$. As shown in Fig.~\ref{fig:1}(d), $Q=\pm 1/3$. The relative azimuthal angle between two sites separated by $\boldsymbol{r}_j-\boldsymbol{r}_i=m_1\boldsymbol{a}_1'+m_2\boldsymbol{a}_2'$ is $2\pi Q(m_1+m_2)$. For example, the direction of the spins in the $xy$ plane on atom $1$ and atom $3$ ($m_1=m_2=1$) differ by $120^\circ$, as indicated by the blue arrows in Fig.~\ref{fig:1}(a). Further comparing to Eq.~\ref{eq:gs}, one finds that for sublattice~A, $\boldsymbol{Q}$ satisfies $Q=\mathrm{sgn}(D)/3$, while for sublattice~B, $Q=-\mathrm{sgn}(D)/3$.
Therefore, the honeycomb lattice can be regarded as two single-$\boldsymbol{Q}$ triangular lattices coupled.

In conclusion, the structure of the phase diagram in Fig.~\ref{fig:1}(b) results from the competition between different terms in the spin Hamiltonian Eq.~\eqref{hami}. 
The ferromagnetic Heisenberg interaction favors parallel spin alignment, the DMI stabilizes an in-plane 120$^\circ$ noncollinear order, 
and the Zeeman term aligns spins along the external field (out of plane).

\begin{figure}
\includegraphics[width=\linewidth]{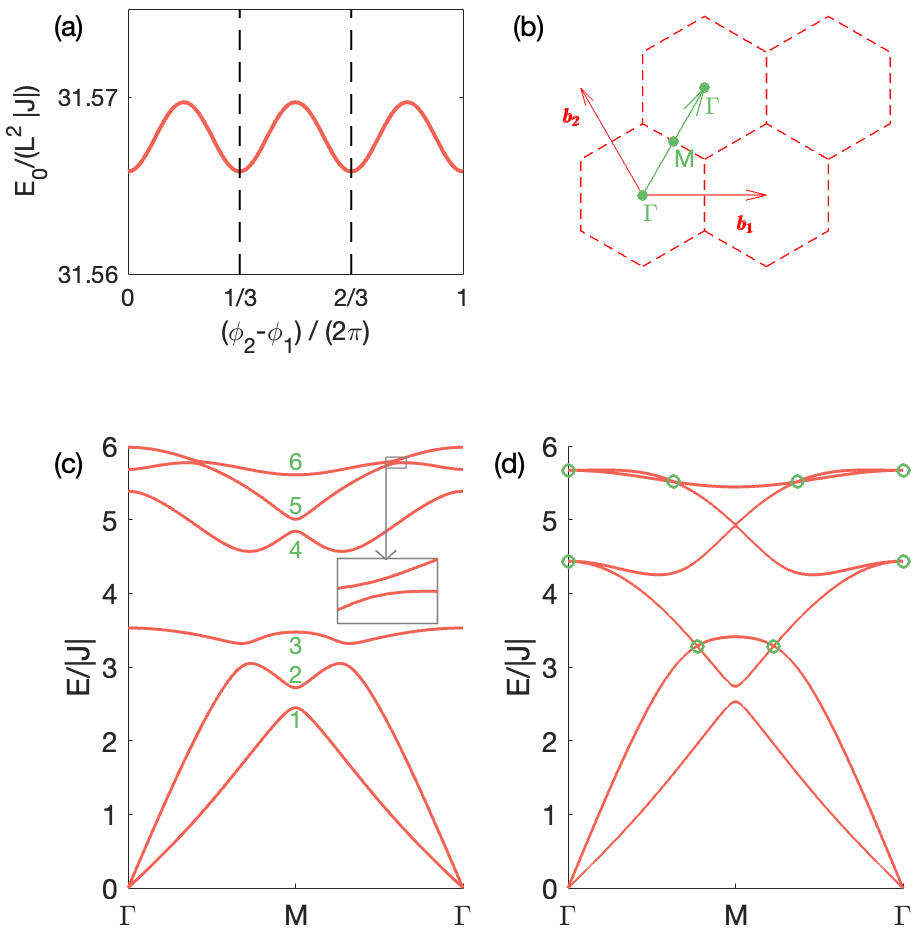}
    \caption{\textbf{Zero-point energy and magnon bands.} \textbf{(a)} Zero-point energy (per unit cell) as a function of $\phi_2-\phi_1$ in the strong-$D$ phase. The parameters used are $D/|J|=0.8$ and $h/|J|=0.3$. \textbf{(b)} High-symmetry points in momentum space used in panels (c,d). \textbf{(c-d)} Magnon bands in the strong-$D$ phase with $\phi_2=\phi_1$: (c) $D/|J|=0.8$, $h/|J|=0.3$. The inset shows the band gap between $E_6$ and $E_5$. The green numbers are band indices. (d) $D/|J|=0.8$, $h=0$. Green circles mark additional band degeneracies compared with panel (c).}
    \label{fig:2}
\end{figure}

\subsection{Linear spin wave theory}
To determine the magnon spectrum of our system, we perform a Holstein-Primakoff (HP) transformation~\cite{PhysRev.58.1098}, \begin{equation}
\begin{aligned}
    \widetilde{S}^{x}_\mathbf{r} \approx\frac{\sqrt{2 S}}{2} (\beta_\mathbf{r}+\beta^{\dagger}_\mathbf{r}),\\ \widetilde{S}^{y}_\mathbf{r} \approx\frac{\sqrt{2 S}}{2 i} (\beta_\mathbf{r}-\beta^{\dagger}_\mathbf{r}),\\ \quad \widetilde{S}^z_\mathbf{r}=-\beta^\dagger_\mathbf{r} \beta_\mathbf{r}+S,
\end{aligned}
    \label{eq:hp}
\end{equation} 
where $\beta^\dagger_\mathbf{r} (\beta_\mathbf{r})$ denotes boson creation (annihilation) operator at position $\mathbf{r}$ and $\widetilde{x}-\widetilde{y}-\widetilde{z}$ denotes a local frame such that $\widetilde{z}$ points along the classical spin vector. Generally speaking, the LSW Hamiltonian is given by
\begin{equation}
H_m= \tfrac{1}{2} \sum_{\boldsymbol{k}}
\begin{bmatrix}
\boldsymbol{\beta}_{\boldsymbol{k}}^{\dagger} \boldsymbol{\beta}_{-\boldsymbol{k}}
\end{bmatrix}
\boldsymbol{H}_{\boldsymbol{k}}
\begin{bmatrix}
\boldsymbol{\beta}_{\boldsymbol{k}} \\
\boldsymbol{\beta}_{-\boldsymbol{k}}^{\dagger}
\end{bmatrix},
\end{equation}
where $\boldsymbol{\beta}_{\boldsymbol{k}}^{\dagger} \equiv 
\left[\beta_{1,\boldsymbol{k}}^{\dagger}, \ldots, \beta_{n,\boldsymbol{k}}^{\dagger}\right]$ 
denotes boson creation operators for the $n$ sublattices ($n=2, 6$ for weak- and strong-$D$ phases, respectively).  Here, $\beta_{i,\boldsymbol{k}}$ is defined by $\beta_{i,\boldsymbol{k}}=\frac{1}{\sqrt{N/n}} \sum_{\mathbf{r}} e^{-i \mathbf{k} \cdot \mathbf{r}} \beta_{\mathbf{r}\in i}$.

In the weak-$D$ phase, the system hosts two magnon bands. The LSW Hamiltonian is 
\begin{widetext}
\begin{equation}
\begin{aligned}
H^{\text{weak}}_m=\sum_{\boldsymbol{k}} \begin{bmatrix}a_{\boldsymbol{k}}^\dagger b_{\boldsymbol{k}}^\dagger\end{bmatrix}\boldsymbol{H}_{\boldsymbol{k}}^{\text{weak}}\begin{bmatrix}
a_{\boldsymbol{k}} \\
b_{\boldsymbol{k}}
\end{bmatrix},\quad
\boldsymbol{H}_{\boldsymbol{k}}^{\text{weak}}=\left(\begin{array}{cc}
-3JS+\mathrm{sgn}(M^z)2DSg_{\boldsymbol{k}}+|h|& JSf_{\boldsymbol{k}} \\
JSf_{\boldsymbol{k}}^* & -3JS-\mathrm{sgn}(M^z)2DSg_{\boldsymbol{k}}+|h|
\end{array} \right),
\end{aligned}
\end{equation}    
\end{widetext}
where $$f_{\boldsymbol{k}}=e^{i\left(\frac{k_1^\prime2\pi}{3}-\frac{k_2^\prime2\pi}{3}\right)}\left(1+e^{ik_2^\prime2\pi}+e^{-ik_1^\prime2\pi}\right),$$ and $$g_{\boldsymbol{k}}=\sin(k_1^\prime2\pi)+\sin(k_2^\prime2\pi)-\sin[(k_1^\prime+k_2^\prime)2\pi].$$ Here, $a_{\boldsymbol{k}}$ and $b_{\boldsymbol{k}}$ denote bosonic operators on sublattices A and B, respectively. The corresponding spin-wave dispersion relation is given by
\begin{equation}
E_\pm(\boldsymbol{k})=-3JS+|h|\pm\sqrt{(2DSg_{\boldsymbol{k}})^2+(JS|f_{\boldsymbol{k}}|)^2}.\label{eq:smallD}
\end{equation} 
The physical requirement $E_\pm(\boldsymbol{k})>0$ imposes $|D|\leqslant D_c$, consistent with the classical ground-state phase boundary. In this weak-$D$ region, a finite value of $D$ opens a band gap at the $\pm K$ point, $(k_1^\prime,k_2^\prime)=\pm(1/3,1/3)$. Appendix~\ref{app.A} provides more details about LSW theory in the weak-$D$ phase.

In the strong-$D$ phase, the system hosts six magnon bands. The LSW Hamiltonian is
\begin{equation}
H_m^{\text{strong}}= \tfrac{1}{2} \sum_{\boldsymbol{k}}
\begin{bmatrix}
\boldsymbol{\beta}_{\boldsymbol{k}}^{\dagger} \boldsymbol{\beta}_{-\boldsymbol{k}}
\end{bmatrix}
\boldsymbol{H}^{\text{strong}}_{\boldsymbol{k}}
\begin{bmatrix}
\boldsymbol{\beta}_{\boldsymbol{k}} \\
\boldsymbol{\beta}_{-\boldsymbol{k}}^{\dagger}
\end{bmatrix},
\end{equation}
where $\boldsymbol{\beta}_{\boldsymbol{k}}^{\dagger}=\left[\beta_{1,\boldsymbol{k}}^{\dagger}, \ldots, \beta_{6,\boldsymbol{k}}^{\dagger}\right]$ (see Appendix~\ref{app.B} for more details). We diagonalize the LSW Hamiltonian using a paraunitary transformation, $\boldsymbol{T}_{\boldsymbol{k}}$, $\boldsymbol{T}_{\boldsymbol{k}}^{\dagger}\,\boldsymbol{H}_{\boldsymbol{k}}\,\boldsymbol{T}_{\boldsymbol{k}}
=
\begin{bmatrix}
\boldsymbol{E}_{\boldsymbol{k}} & 0 \\
0 & \boldsymbol{E}_{-\boldsymbol{k}}
\end{bmatrix}$~\cite{PhysRevB.87.174427},
and obtain
\begin{equation}
H_m^{\text{strong}}= \sum_{\boldsymbol{k},\alpha} E_{\alpha}(\boldsymbol{k})\!\left(n_{\boldsymbol{k},\alpha} + \tfrac{1}{2}\right),
\end{equation} 
where $n_{\boldsymbol{k},\alpha}$ is the number operator of magnon excitation and $\alpha=1,\ldots,6$ denotes the band index, with the lowest band labeled as $\alpha = 1$. The zero-point energy,  
\begin{equation}
    E_0=\tfrac{1}{2}\sum_{\boldsymbol{k},\alpha}E_{\alpha}(\boldsymbol{k}),
\end{equation}
depends on the phase difference $\phi_2-\phi_1$. Figure~\ref{fig:2}(a) displays the areal density of the zero-point energy, $E_0/L^2$, where $L$ is the size of the system, as a function of $\phi_2 - \phi_1$. Minimizing the energy density selects $\phi_2-\phi_1=0,,2\pi/3,$ or $4\pi/3$, corresponding to degenerate topological magnon bands. This effect is known as quantum order-by-disorder mechanism~\cite{PhysRevB.102.220405}.
In the following calculations, we set $\phi_2 = \phi_1$. Choosing the same point as marked in the phase diagram Fig.~\ref{fig:1} (b), we plot the magnon bands in Fig~\ref{fig:2}(c) along the path shown in Fig.~\ref{fig:2}(b). The emergence of band gaps is notable compared to the case where $h=0$ leads to band degeneracies, as circled in Fig~\ref{fig:2}(d). The study of their topological properties is naturally motivated in the following sections. \color{black}

\section{Band topology and thermal Hall effect}\label{sec:CN}

\begin{figure*}[ht!]
\includegraphics[width=\linewidth]{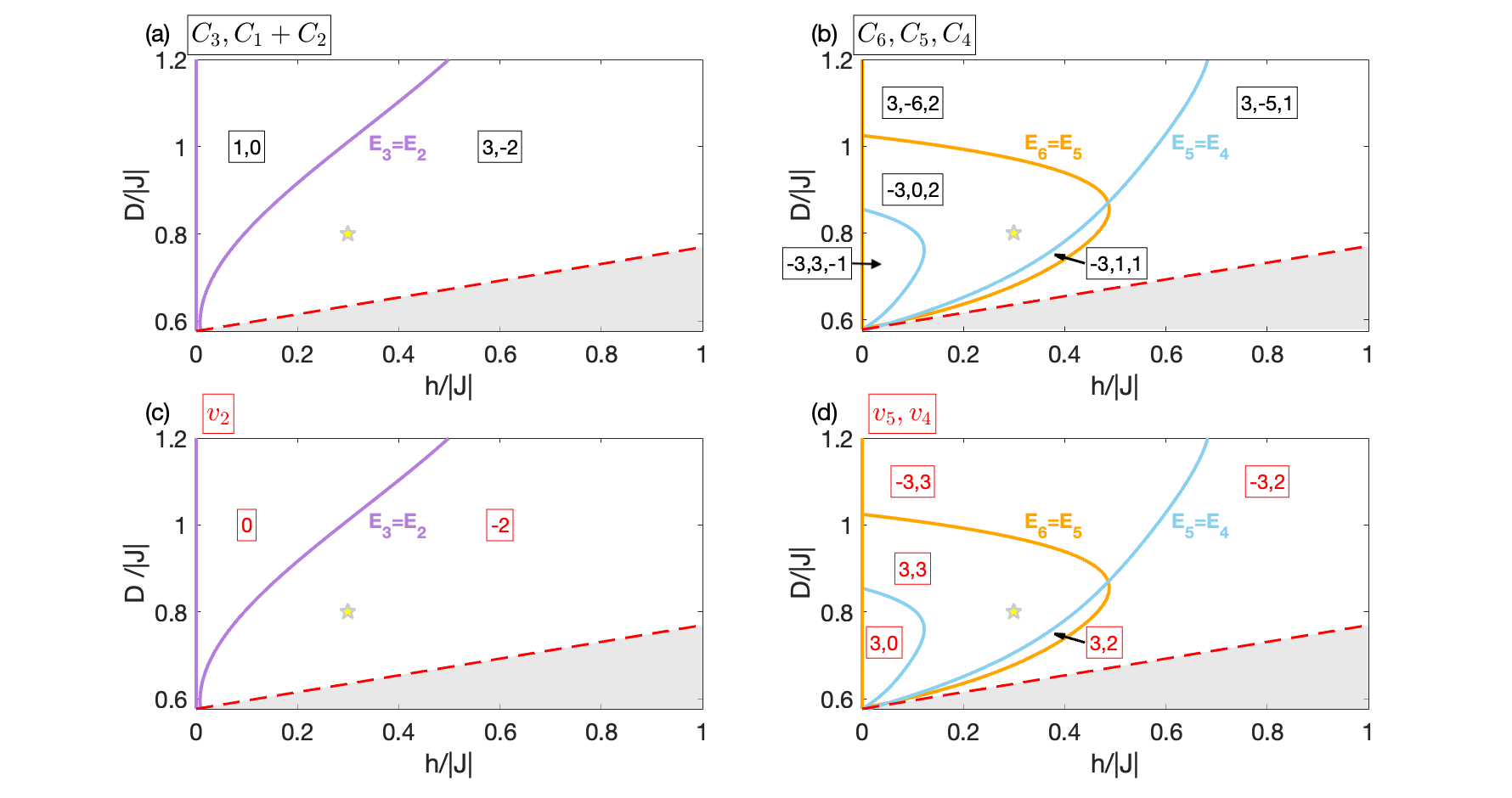}
    \caption{\textbf{Topological phase diagram.} 
% \textbf{(a)} $C_1=v_1$ in the weak-$D$ phase (gray shaded region) and $\sum_{i=1}^3 C_i=v_3$ in the strong-$D$ phase (white background). 
\textbf{(a, b)} Topological characteristics in the phase diagram for the strong-$D$ phase: (a) $C_3$ and $C_2+C_1$; (b) $C_6$, $C_5$, and $C_4$. \textbf{(c, d)} Chiral edge states numbers in the strong-$D$ phase: (c) $v_2$; (d) $v_5$ and $v_4$. The yellow stars indicate the parameters $D/|J| = 0.8$ and $h/|J| = 0.3$ used in Fig.~\ref{fig:2}(c).}
    \label{fig:3}
\end{figure*}

To study the band topology, we calculate the Chern numbers of the magnon bands, defined as 
\begin{equation}
    C_\alpha = -\frac{1}{2\pi} \int_{\mathrm{BZ}} 
    B_\alpha(\boldsymbol{k})\, d^2\boldsymbol{k},
\end{equation}
where $\alpha$ is the band index, the Berry curvature is given by $B_\alpha(\boldsymbol{k})= \big(\nabla \times \boldsymbol{A}_\alpha(\boldsymbol{k})\big)^z$, $A_{\alpha}^v(\boldsymbol{k}) \equiv-\operatorname{Im} \left[\boldsymbol{\sigma}_3 \boldsymbol{T}_{\boldsymbol{k}}^{\dagger} \boldsymbol{\sigma}_3\left(\partial_{k_v} \boldsymbol{T}_{\boldsymbol{k}}\right)\right]_{\alpha,\alpha}$, and $\boldsymbol{\sigma}_3=\left(\begin{array}{cc}1_{n \times n} & 0 \\ 0 & -1_{n \times n}  \end{array}\right)$ with $n$ being the number of atoms in the unit cell~\cite{PhysRevB.87.174427}.

In the weak-$D$ phase, $D\neq 0$ opens a band gap 
$\Delta E = 6\sqrt{3}|D|S$. Since $\boldsymbol{H}_{\boldsymbol{k}}^{\text{weak}}\neq (\boldsymbol{H}_{\boldsymbol{-k}}^{\text{weak}})^*$ for $D\neq 0$, nonzero Chern numbers are allowed. 
The Chern number of the lowest band is  
\begin{equation}
C_1 = \mathrm{sgn}(S^z)\,\mathrm{sgn}(D).
\end{equation}
The DMI, which characterizes the system's chirality, primarily determines $\mathrm{sgn}(C_1)$, whereas the Zeeman term controls $\mathrm{sgn}(S^z)$ and thus also affects $\mathrm{sgn}(C_1)$~\cite{Owerre_2016,PhysRevB.90.024412}.

Unlike the collinear weak-$D$ phase, the strong-$D$ phase exhibits noncollinear and noncoplanar spin structures, leading to more magnon bands and richer topological phases with larger Chern numbers. In this phase, both $D$ and $h$ affect the band gaps. When $h=0$, the system is symmetric under the combined operation of time reversal and a $180^\circ$ spin rotation around the $z$-axis, $\tilde{T}=\exp(-i\pi S^z)T$. 
This symmetry enforces $\boldsymbol{H}_{\boldsymbol{k}}^{\text{strong}}=(\boldsymbol{H}_{-\boldsymbol{k}}^{\text{strong}})^*$, resulting in topologically trivial magnon bands~\cite{PhysRevB.100.064412}. 
By contrast, a finite $h$ induces a noncoplanar spin configuration with finite scalar chirality,  
\begin{equation}
    \chi = \boldsymbol{S}_{1}\cdot(\boldsymbol{S}_{3}\times \boldsymbol{S}_{5}) 
     = \boldsymbol{S}_{2}\cdot(\boldsymbol{S}_{4}\times \boldsymbol{S}_{6}),
\end{equation}
which changes sign upon reversing $\mathrm{sgn}(D)$. In this case, the $\tilde{T}$ symmetry is broken and $\boldsymbol{H}_{\boldsymbol{k}}^{\text{strong}}\neq (\boldsymbol{H}_{-\boldsymbol{k}}^{\text{strong}})^*$, allowing for nonzero Chern numbers. In particular, the sum of Chern numbers of the lowest three bands always satisfy 
\begin{equation}
\sum_{\alpha=1}^3 C_\alpha = \mathrm{sgn}(S^z)\,\mathrm{sgn}(D).
\end{equation}
Since $\sum_{\alpha=1}^6 C_\alpha=0$~\cite{asboth2016short}, we also have 
\begin{equation}
    \sum_{\alpha=4}^6 C_\alpha=-\mathrm{sgn}(S^z)\,\mathrm{sgn}(D).
\end{equation}
Reversing the sign of either $D$ or $h$ flips all Chern numbers without altering the energy spectrum. Hence $h=0$ marks a topological phase transition where the magnon bands become degenerate, as shown in Fig.~\ref{fig:2}(d).

\begin{figure*}[ht!]
\includegraphics[width=\linewidth]{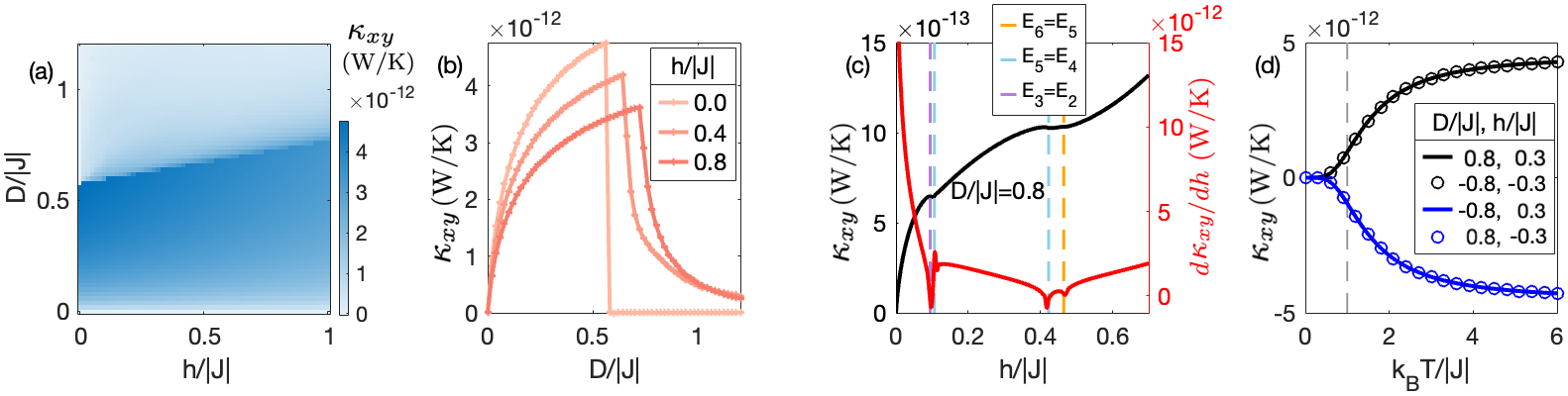}
\caption{\textbf{Magnon Thermal Hall conductivity.} \textbf{(a)} $\kappa_{xy}$ as a function of $D$ and $h$. \textbf{(b)} $\kappa_{xy}$ versus $D$ for selected values of $h$. \textbf{(c)} $\kappa_{xy}$ and $d\kappa_{xy}/dh$ at $D/|J|=0.8$ versus $h$. The dashed lines indicate the parameters at which topological phase transitions occur, as shown in Fig.~\ref{fig:2}(b,c). \textbf{(d)} Temperature dependence of $\kappa_{xy}$ with different values of $D$ and $h$. The dashed line indicates $T=J/k_B$. Panels (a-c) are calculated at $T=J/k_B$.}
    \label{fig:4}
\end{figure*}

Within each band group, namely the upper three and the lower three, there exists a cascade of topological phase transitions characterized by gap opening/closing. For the lower three bands, since $E_1$ and $E_2$ are always gapless at $\Gamma$ point, as shown in Figs.~\ref{fig:2}(c)(d), the individual Chern numbers $C_1$ and $C_2$ are ill-defined. We hence only consider their sum $C_1+C_2$. Energy degeneracy, $E_3=E_2$, changes $C_3$ and $C_1+C_2$, as shown in Fig.~\ref{fig:3}(a). For the upper three bands, the closing of the band gap between the sixth and fifth band, namely $E_6=E_5$, leads to a change in $C_6$ and $C_5$; 
$E_5=E_4$ changes $C_5$ and $C_4$, as shown in Fig.~\ref{fig:3}(b). For a typical phase, as marked by the yellow star, corresponding gapped topological magnon bands shown in Fig~\ref{fig:2}(c), the Chern numbers are determined as: $C_6=-3$, $C_5=0$, $C_4=2$, $C_3=3$, and $C_1+C_2=-2$. More details about band gaps can be found in Appendix~\ref{app.B}.

The number of chiral edge states within the gap between the $m$-th and $(m+1)$-th bands is given by  
\begin{equation}
v_m \equiv \sum_{j=1}^m C_j,\label{eq:edge}
\end{equation}
where $\mathrm{sgn}(v_m)=+1$ ($-1$) corresponds to counterclockwise (clockwise) chiral edge modes~\cite{PhysRevB.87.174427}. 
While the weak-$D$ phase is characterized by $v_1=C_1$, the strong-$D$ phase features $v_3=\sum_{\alpha=1}^3 C_\alpha$. 
Moreover, the strong-$D$ phase can host larger $|v_m|$ values, such as $2$ and $3$. We plot $v_2$, $v_4$, and $v_5$ as functions of $D$ and $h$ in Figs.~\ref{fig:3}(c,d). Figure.~\ref{fig:2}(c) represents an example with $v_5=3$ $v_4=3$, $v_3=1$ and $v_2$=-2. 

It is well established that a nonzero Berry curvature gives rise to a finite anomalous thermal Hall conductivity,  
\begin{equation}
    \kappa_{xy} = -\frac{k_B^2 T}{\hbar (2\pi)^2}\sum_{\alpha} \int c_2(\rho_{\alpha, \boldsymbol{k}})\,B_{\alpha}(\boldsymbol{k}) d^2\boldsymbol{k},
\end{equation}
 where $c_2(\rho)=(1+ \rho)\left(\log \frac{1+\rho}{\rho}\right)^2-(\log \rho)^2-2 \mathrm{Li}_2(-\rho)$ with $\mathrm{Li}_2(z)$ the polylogarithm function and the Bose-Einstein distribution $\rho_{\alpha, \boldsymbol{k}}=1 /\left(\exp \left(E_{\alpha}( \boldsymbol{k}) / k_B T\right)-1\right)$~\cite{PhysRevLett.106.197202}. As shown in Fig.~\ref{fig:4}(a), $\kappa_{xy}$ is strongly influenced by the values of $D$ and $h$, and presents clear signatures of the phase transition from the weak- to strong-$D$ regime. For a given $h$, $\kappa_{xy}$ varies nonmonotonically with $D$ [Fig.~\ref{fig:4}(b)]. Since $\frac{d c_2(\rho)}{d\rho} 
= \left(\log\frac{1+\rho}{\rho}\right)^2 \geqslant 0$, $c_2(\rho)$ increases with increasing $\rho$, thus in turn with decreasing energy $E_\alpha$. In the weak-$D$ phase, although the upper and lower bands carry opposite Berry curvatures, $B_1(\boldsymbol{k})=-B_2(\boldsymbol{k})<0$, the lower band $E_1(\boldsymbol{k})$ dominates the contribution, yielding a positive $\kappa_{xy}$~\cite{Owerre20162}. Moreover, as $D$ increases, the band gap widens (see Fig.~\ref{fig:S1}(a) in Appendix~\ref{app.A}), and the contribution from the lower band becomes more dominant, resulting in an overall increase of $\kappa_{xy}$ with $D$.
In the strong-$D$ phase, the lower (upper) three bands contribute to positive (negative) Berry curvature: $\sum_{\alpha=1}^3 B_\alpha(\boldsymbol{k}) = -\sum_{\alpha=4}^6 B_\alpha(\boldsymbol{k}) < 0$ and larger population of the former group leads to positive $\kappa_{xy}$. However, the gap between two groups (namely $E_4-E_3$) decreases as $D$ increases for a given $h$, as shown in Fig.~\ref{fig:S3}(c), (see Appendix~\ref{app.B} for more details). This qualitatively explains why $\kappa_{xy}$ decreases with increasing $D$ in the strong-$D$ phase. It can also be found, in Fig.~\ref{fig:S3}(c), that this gap increases with increasing $h$ at a given $D$ value. $\kappa_{xy}$ thus increase with $h$, as shown by the black curve in Fig.~\ref{fig:4}(c). In addition, $\kappa_{xy}$ exhibits clear features related to topological phase transitions~\cite{PhysRevB.100.064412}: its first derivative (red curve) with respect to the Zeeman field $h$ diverges at topological transition points, where band degeneracies occur. Furthermore, the divergence at $h = 0$ corresponds to the sign reversal of Berry curvatures. To further illustrate this behavior, we show in Fig.~\ref{fig:4}(d) the temperature dependence of $\kappa_{xy}$ for different parameter sets. Reversing the sign of $h$ or $D$ transforms $\kappa_{xy} \to -\kappa_{xy}$, consistent with the characteristics of distinct topological phases.
 
 In summary, the strong-$D$ phase presents richer topological phases, which can lead to more chiral edge states compared to weak-$D$ phase and anomalous thermal Hall effect.

\section{Magnon-phonon coupling and hybridized bands}

To further examine the difference between the weak- and strong-$D$ phases, we analyze the magnon-phonon coupling. The phonon Hamiltonian is given by
\begin{equation}
\begin{aligned}
    H_{p}^{\text{total}}&=\sum_i \frac{(\boldsymbol{p}_i)^2}{2 M}+\frac{1}{2} K_1 \sum_{\langle i, j\rangle} \left[\left(\boldsymbol{u}_i-\boldsymbol{u}_j\right) \cdot \boldsymbol{d}_{i j}\right]^2\\&+\frac{1}{2} K_2 \sum_{\langle\langle i, j\rangle \rangle} \left[\left(\boldsymbol{u}_i-\boldsymbol{u}_j\right) \cdot\boldsymbol{d}_{i j}\right]^2,
\end{aligned}
\end{equation}
where $\boldsymbol{d}_{ij}=(d_{ij}^x,d_{ij}^y)$ is the unit vector from site $j$ to site $i$, while $\boldsymbol{p}_i=(p_i^x,p_i^y)$ and $\boldsymbol{u}_i=(u_i^x,u_i^y)$ denote the momentum and displacement operators, $K_1$ and $K_2$ are the NN and the NNN phonon spring constants, respectively~\cite{PhysRevB.104.045139}. For simplicity, we consider lattice vibrations along the $x$ direction, the Hamiltonian reads,
\begin{equation}
\begin{aligned}
    H_{p}&=\sum_i \frac{(p_i^x)^2}{2 M}+\frac{1}{2} K_1 \sum_{\langle i, j\rangle} \left[\left(u_i^x-u_j^x\right) d_{i j}^x\right]^2\\&+\frac{1}{2} K_2 \sum_{\langle\langle i, j\rangle \rangle} \left[\left(u_i^x-u_j^x\right) d_{i j}^x\right]^2,
\end{aligned}
\end{equation}
where the phonon displacement and momentum operators can be expressed 
in terms of bosonic operators as  

\begin{equation}
    \begin{aligned}
        u_i^x&=\tfrac{1}{\sqrt{2}}\left(\alpha_i+\alpha_i^\dagger\right)\\
        p_i^x&=\tfrac{1}{\sqrt{2}}\left(-\mathrm{i}\alpha_i+\mathrm{i}\alpha_i^\dagger\right)
    \end{aligned}\label{eq:pho_bos}
\end{equation}

We can now assume that the lattice displacements are small and expand the DMI and Heisenberg coupling around the equilibrium coordinates~\cite{PhysRevB.106.214424} as
\begin{equation}
\begin{aligned}
    D_{i j} &\approx \bar{D}_{i j}+\partial D_{i j} / \partial r_{i j}\left[\left(u_i^x-u_j^x\right)  d_{i j}^x\right]\\
    J_{i j} &\approx \bar{J}_{i j}+\partial J_{i j} / \partial r_{i j}\left[\left(u_i^x-u_j^x\right)  d_{i j}^x\right]
\end{aligned}
\end{equation}
where higher-order terms are neglected.
This naturally induces a magnon-phonon coupling that can be rewritten as,
\begin{equation}
\begin{aligned}
    H_{mp}&=\sum_{\langle ij\rangle}\partial J_{i j} / \partial r_{i j}\left[\left(u_i^x-u_j^x\right)  d_{i j}^x\right]\vec{S}_i\cdot\vec{S}_j\\&+\sum_{\langle\langle ij\rangle\rangle}\partial D_{i j} / \partial r_{i j}\left[\left(u_i^x-u_j^x\right)  d_{i j}^x\right] (S_i^xS_j^y-S_i^yS_j^x).
\end{aligned}\label{eq:mp}
\end{equation}

We introduce a local frame at each site before performing the HP transformation, 
\begin{equation}
\begin{aligned}
    & \left(\begin{array}{c}
S_i^x \\
S_i^y \\
S_i^z
\end{array}\right)=R_i\left(\begin{array}{c}
\tilde{S}_i^x \\
\tilde{S}_i^y  \\
\tilde{S}_i^z
\end{array}\right)\\
&R_i=\left(\begin{array}{ccc}
    -\sin\phi_i & -\cos\theta_i \cos\phi_i & \sin\theta_i \cos\phi_i \\
    \cos\phi_i & -\cos\theta_i \sin\phi_i & \sin\theta_i \sin\phi_i \\
    0 & \sin\theta_i & \cos\theta_i
    \end{array}\right).
\end{aligned}\label{eq:frame}
\end{equation}
We set $\phi_i=-\frac{\pi}{2}$ and $0$ for $\theta_i=0$ and $\pi$, respectively. From Eq.~\ref{eq:hp} about HP transformation and Eq.~\ref{eq:pho_bos} about bosonic operators for phonon, we know that only the terms $\tilde{S}_i^x \tilde{S}_j^z$, $\tilde{S}_i^z \tilde{S}_j^x$, $\tilde{S}_i^y \tilde{S}_j^z$, and $\tilde{S}_i^z \tilde{S}_j^y$ with $\tilde{S}_i^z=\tilde{S}_j^z \approx S$ contribute to the quadratic terms in Eq.~\ref{eq:mp}. With uniform $S^z$, the relevant contributions from $S_i^x S_j^y - S_i^y S_j^x$ and $\vec{S}_i\cdot\vec{S}_j$ are
\begin{widetext}
\begin{equation}
\begin{aligned}
        S_i^xS_j^y-S_i^yS_j^x &\rightarrow-\sin(\theta)\cos(\Delta\phi)(\tilde{S}_i^x \tilde{S}_j^z-\tilde{S}_i^z \tilde{S}_j^x)+\cos(\theta)\sin(\theta)\sin(\Delta\phi)(\tilde{S}_i^y \tilde{S}_j^z+\tilde{S}_i^z \tilde{S}_j^y)\\
        \vec{S}_i\cdot\vec{S}_j&\rightarrow 
        -\sin(\theta)\sin(\Delta\phi)(\tilde{S}_i^x \tilde{S}_j^z-\tilde{S}_i^z \tilde{S}_j^x)+\cos(\theta)\sin(\theta)(1-\cos(\Delta\phi))(\tilde{S}_i^y \tilde{S}_j^z+\tilde{S}_i^z \tilde{S}_j^y),
\end{aligned}
\end{equation}
\end{widetext}
where $\Delta\phi=\phi_i-\phi_j$. These terms are nonzero only when 
\begin{equation}
    \theta \neq 0 \text{ and }  \phi_i\neq\phi_j.
\end{equation}
Therefore, the magnon-phonon coupling vanishes in the weak-$D$ phase where spins are collinear, but it is finite in the strong-$D$ phase. In other words, the non-collinear spin configuration naturally allows for a quadratic magnon-phonon coupling.

\begin{figure}
\includegraphics[width=\linewidth]{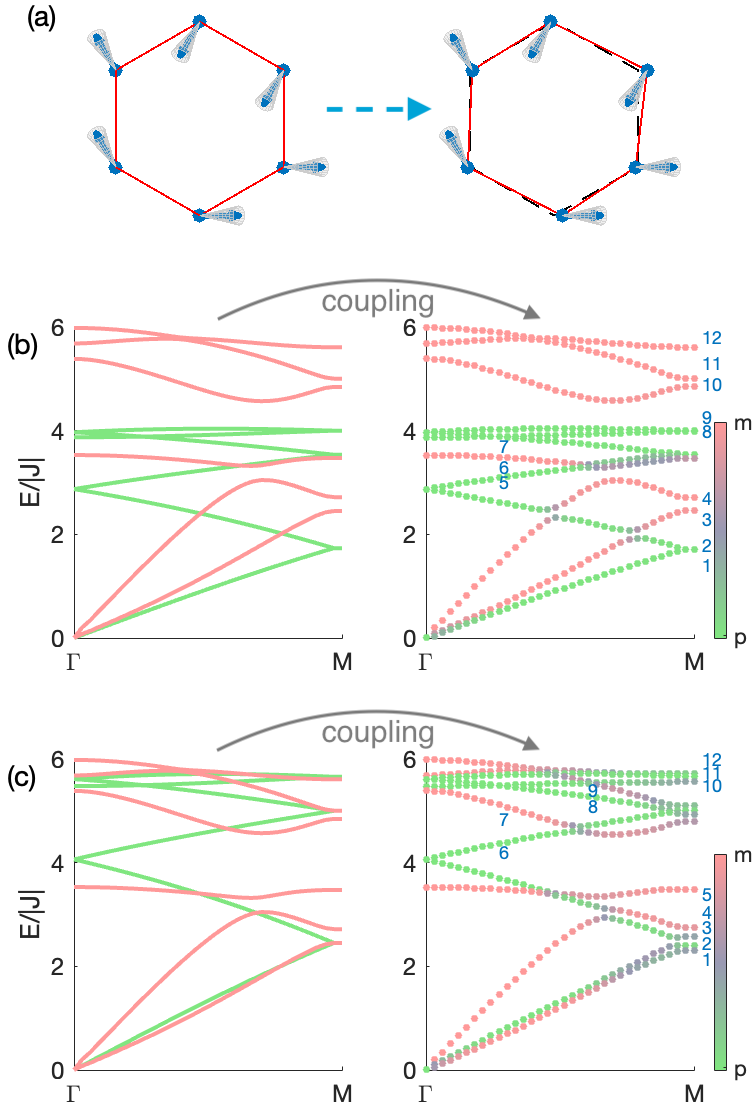}
 \caption{\textbf{Magnon-phonon coupling.} 
    \textbf{(a)} Schematic illustration of magnon-phonon hybrid excitations. \textbf{(b)} Decoupled and coupled bands for $K_1/|J|=5$, $K_2/|J|=1$, and $\partial D_{ij}/\partial r_{ij}=0.3$. \textbf{(c)} Decoupled and coupled magnon-phonon bands for $K_1/|J|=10$, $K_2/|J|=2$, and $\partial D_{ij}/\partial r_{ij}=0.4$. Panels (b-c) are calculated using $M=1$, $D/|J|=0.8$, $h/|J|=0.3$ and $\partial J_{ij}/\partial r_{ij}=0$. In the decoupled case, the pink and green lines denote the magnon and phonon bands, respectively. When the coupling is turned on, the color represents the relative weights of the spin and lattice vibrations: pink (green) indicates a predominantly spin (lattice) character, while gray denotes strong magnon-phonon hybridization.}
    \label{fig:5}
\end{figure}

In the strong-$D$ phase, including the magnon-phonon coupling, the total Hamiltonian is $H=E_{GS}+H_m^{\text{strong}}+H_{p}+H_{mp}$, where $E_{GS}$ is classical ground state energy. We assume that the coupling is weak, the original ground state is not changed in the presence of magnon-phonon coupling~\cite{PhysRevB.106.214424,PhysRevLett.117.207203}. Figure~\ref{fig:5}(a) shows a schematic illustration of the magnon-phonon coupling. We consider magnons accompanied by lattice vibrations, which modulate the interatomic distances and hence the coupling strength, as formulated above. We obtain the quadratic bosonic Hamiltonian
\begin{equation}
    H_m^{\text{strong}}+H_p+H_{mp}= \tfrac{1}{2} \sum_{\boldsymbol{k}}
\begin{bmatrix}
\boldsymbol{\alpha}_{\boldsymbol{k}}^{\dagger}
\boldsymbol{\beta}_{\boldsymbol{k}}^{\dagger} 
\boldsymbol{\alpha}_{-\boldsymbol{k}}
\boldsymbol{\beta}_{-\boldsymbol{k}}
\end{bmatrix}
\boldsymbol{H}_{\boldsymbol{k},mp}
\begin{bmatrix}
\boldsymbol{\alpha}_{\boldsymbol{k}}\\
\boldsymbol{\beta}_{\boldsymbol{k}} \\
\boldsymbol{\alpha}_{-\boldsymbol{k}}^{\dagger}\\
\boldsymbol{\beta}_{-\boldsymbol{k}}^{\dagger}
\end{bmatrix}.\label{eq:Hmp}
\end{equation}
Diagonalizing $\boldsymbol{H}_{\boldsymbol{k},mp}$ yields the magnon-phonon hybrid bands, and the magnon-phonon weight ratio can be extracted from the corresponding eigenvectors. 

Figures~\ref{fig:5} (b) and (c) present two cases of coupled magnon-phonon bands, respectively, with corresponding decoupled bands as references. Without hybridization, the magnon bands (pink) are topological (discussed before in Sec.~\ref{sec:CN}) while the phonon bands (green) are topologically trivial. With magnon-phonon coupling, the color represents the relative weights of the spin and lattice vibrations,
\begin{equation}
    w=\frac{\sum_{j=1}^6|m_j|^2}{\sum_{j=1}^6|m_j|^2+\sum_{i=1}^6|p_i|^2}.
\end{equation} 
where $p_i$ and $m_j$ are the phonon and magnon coefficients, respectively, in the eigenvectors of the coupled Hamiltonian $\boldsymbol{H}_{\boldsymbol{k},mp}$: 
\begin{equation}
\gamma_{\boldsymbol{k}}=\sum_{i=1}^6 p_i \alpha_{i,\boldsymbol{k}}+\sum_{j=1}^6 m_j \beta_{j,\boldsymbol{k}}.
\end{equation} 
$w=1$ ($w=0$) corresponds to purely spin (lattice) character, shown in pink (green), while $w=0.5$ indicates strong hybridization, shown in gray. Such strong hybridization occurs when the magnon and phonon frequencies are close to each other. This coupling opens a gap between the magnon and phonon bands, which not only serves as an experimental signature but also gives rise to topological hybrid bands~\cite{PhysRevLett.117.207203,2023Baosong}. For example, as shown in Fig.~\ref{fig:5}(b), when the phonon mode hybridizes with the lower three magnon bands, the resulting Chern numbers are $\sum_{i=1}^3 C_i = 1$, $\sum_{i=4}^9 C_i = 0$, $C_{10} = 2$, $C_{11} = 0$, and $C_{12} = -3$. On the other hand, when the phonon dispersions reach those of the upper three magnon bands [Fig.~\ref{fig:5}(c)], the phonon-magnon coupling is enhanced. In this other example, the Chern numbers change to $\sum_{i=1}^3 C_i = -1$, $\sum_{i=4}^6 C_i = 1$, and $\sum_{i=7}^{12} C_i = 0$.
The listed Chern numbers correspond to the parameters used in the figure and may vary for different parameter choices.

It is worth emphasizing that hybrid magnon-phonon bands only apply to strong-$D$ phase where spins are non-collinear. Within the framework of LSW theory, the quadratic magnon-phonon terms, as formulated in Eq.~\ref{eq:Hmp}, lead to coupling that modifies the bands. Higher order terms, namely the cubic terms containing odd-numbered bosonic operators, will contribute to the magnon-phonon scattering which goes beyond the scope of this work. In the collinear structure of the weak-$D$ phase, the quadratic terms, which originate from $\tilde{S}_i^x \tilde{S}_j^z$, $\tilde{S}_i^z \tilde{S}_j^x$, $\tilde{S}_i^y \tilde{S}_j^z$, and $\tilde{S}_i^z \tilde{S}_j^y$, all vanish, and only the cubic terms remain~\cite{PhysRevB.106.214424}. Going beyond the LSW theory by including more terms in the HP transformation does not change the conclusion. Though quartic magnon-magnon terms can also affect bands within a mean-field treatment~\cite{PhysRevB.111.094430}, quartic magnon-phonon coupling still stems from non-collinear spin interaction.

In summary, the quadratic magnon-phonon coupling, originating from the DMI and Heisenberg interactions, vanishes in the weak-$D$ phase with a ferromagnetic ground state, but emerges in the strong-$D$ phase characterized by a noncollinear and noncoplanar spin configuration. The magnon-phonon hybridization in the strong-$D$ phase mainly takes place near the band crossings of magnon and phonon branches, where the coupling opens gaps and leads to topologically nontrivial hybrid bands.

\section{Conclusion}
In this work, we have studied in detail a ferromagnetic honeycomb magnet with NNN DMI under an external Zeeman field. The model exhibits two distinct ground states characterized by different symmetries and corresponding to the limit of weak and strong DMI respectively. Within the phase diagram, the system undergoes a transition from a collinear ferromagnet (weak-$D$) to a $120^\circ$ ordered state (strong-$D$) that is sharp in absence of Zeeman field and becomes a continuous crossover in presence of the latter. The critical line can be obtained analytically and matches well with the numerical analysis. 
Using LSW theory, we demonstrate that the weak-$D$ phase hosts two magnon topological bands with a DMI-induced gap and non-trivial Chern number $C_1=\mathrm{sgn}(S^z)\mathrm{sgn}(D)$, whereas the strong-$D$ phase exhibits six magnon bands with rich topological structures controlled by both $D$ and $h$. 
In particular, a finite field in the strong-$D$ regime cause noncoplanar spins and enforces $\sum_{\alpha=1}^3 C_\alpha=\mathrm{sgn}(S^z)\mathrm{sgn}(D)$ and other topological phases, yielding up to three chiral edge-state channels. Importantly, we show that these topological phase transitions can be probed using the thermal Hall conductivity.

Finally, we identified a quadratic magnon-phonon coupling that comes from spin-spin interactions can exist in the strong-$D$ phase but vanishes in the weak-$D$ phase. This interaction emerges in the strong-$D$ phase due to noncollinear spin textures and results in magnon-phonon hybrid topological bands.

Overall, our results demonstrate the crucial role of strong DMI in changing both the magnetic ground state and the topological band structure. 
Compared with the weak-$D$ phase, the strong-$D$ phase not only hosts richer topological phases with multiple edge channels but also enables hybrid magnon-phonon modes with nontrivial topology. 
Since the strength of DMI can be experimentally tuned~\cite{PhysRevB.90.020402,Srivastava2018,PhysRevB.106.L100407,PhysRevB.108.134405}, our predictions provide concrete guidelines for material design and experimental exploration in the field of topological magnon spintronics. 

\section*{Acknowledgments}
We thank Professor Ka Shen for helpful discussions. This work was supported by the National Natural Science Foundation of China (Grant No. 12374101 and No. 22Z990204371).

% \nocite{apsrev42Control}
% \bibliographystyle{apsrev4-2} 
% \bibliography{refs}

%apsrev4-2.bst 2019-01-14 (MD) hand-edited version of apsrev4-1.bst
%Control: key (0)
%Control: author (72) initials jnrlst
%Control: editor formatted (1) identically to author
%Control: production of article title (-1) disabled
%Control: page (0) single
%Control: year (1) truncated
%Control: production of eprint (0) enabled
%

\appendix
\section{LSW theory in the weak-$D$ phase}
\label{app.A}
\begin{figure}
\includegraphics[width=\linewidth]{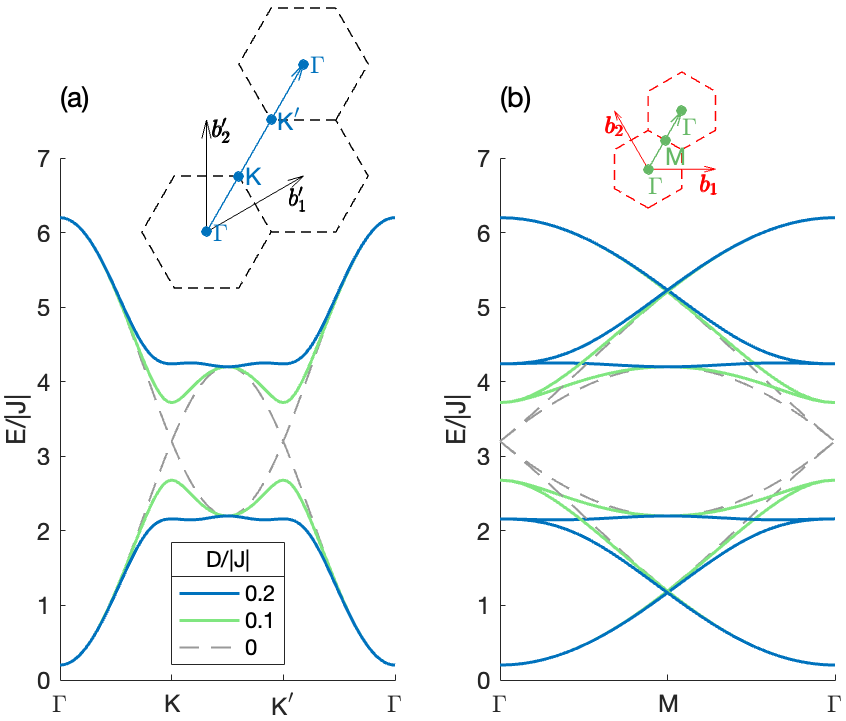}
\caption{\textbf{Weak-$D$ magnon spectrum.}
\textbf{(a)} Spectrum calculated using a unit cell with two atoms for selected values of $D/|J|$. 
\textbf{(b)} Spectrum calculated using a unit cell with six atoms for selected values of $D/|J|$. Panels (a,b) use $h/|J| = 0.2$. The $x$-axis labels correspond to the high-symmetry points in momentum space, as shown in insets.}
    \label{fig:S1}
\end{figure}

In the weak-$D$ phase, the unit cell consists of two sublattices, as shown in Fig.~\ref{fig:1}(a).
Corresponding magnon spectrum is plotted in Fig.~\ref{fig:S1}(a). A finite value of the DMI coupling $D$ opens a band gap at $K$ points, $(k_1^\prime,k_2^\prime)=\pm\left(1/3,1/3\right)$, and enables nonzero Chern numbers. To ensure physical stability, the magnon excitation energy must be non-negative. Since $|g_{\boldsymbol{k}}|$ in Eq.~\ref{eq:smallD} reaches its maximum value $\tfrac{3\sqrt{3}}{2}$ at the $K$ points, one finds that the lower band energy
$E_-(K)=-3JS-|D|S3\sqrt{3}+|h|\geq0$, which imposes the constraint $|D|\leq D_c$. It is equivalent to the requirement of having a ferromagnetic ground state.

\begin{figure}
\includegraphics[width=\linewidth]{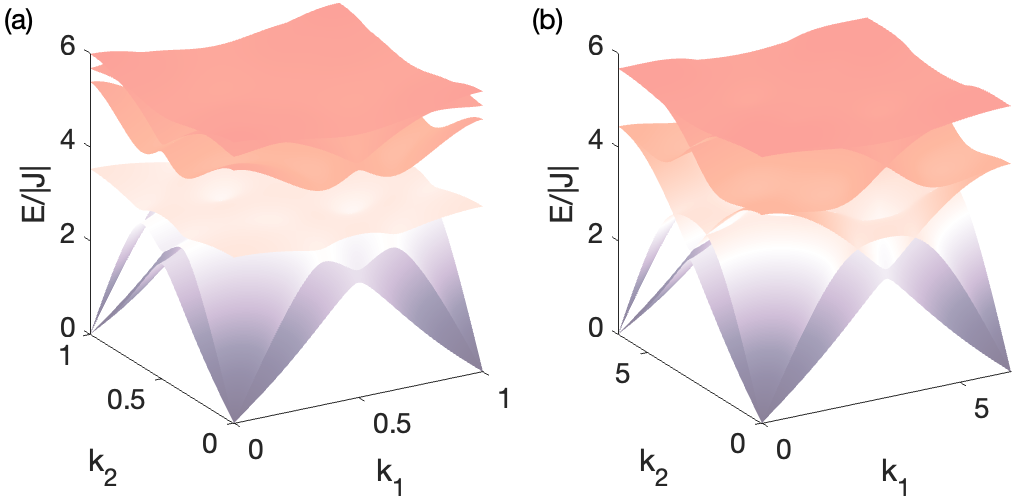}
    \caption{\textbf{Magnon spectrum in the strong-$D$ phase throughout the Brillouin zone.}
\textbf{(a)} $D/|J|=0.8$, $h/|J|=0.3$. \textbf{(b)} $D/|J|=0.8$, $h/|J|=0$.}
    \label{fig:S2}
\end{figure}

To compare with the strong-$D$ phase, we also calculate the weak-$D$ magnon spectrum using a larger unit cell containing six sublattices. The results are shown in Fig.~\ref{fig:S1}(b). In this case, the upper three bands and the lower three bands are each gapless, but there exists a finite gap between the two groups. By contrast, as illustrated in Fig.~\ref{fig:2}, the strong-$D$ phase exhibits more band gaps and thereby supports a richer variety of topological phases.
\section{LSW theory in the strong-$D$ phase}
\label{app.B}

\begin{figure}
\includegraphics[width=\linewidth]{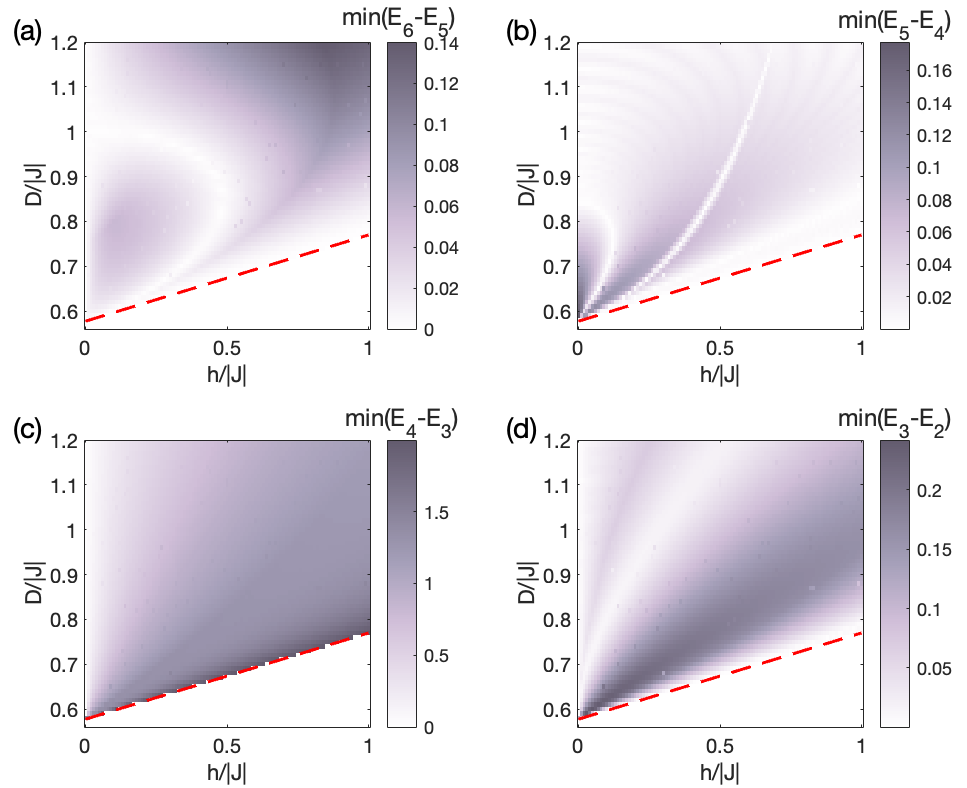}
    \caption{\textbf{Band gaps in the strong-$D$ phase.}
Minimum energy difference between adjacent bands as a function of $D$ and $h$: 
\textbf{(a)} min($E_6-E_5$), 
\textbf{(b)} min($E_5-E_4$), 
\textbf{(c)} min($E_4-E_3$), and 
\textbf{(d)} min($E_3-E_2$). The dashed lines indicating the phase boundary, $D_c$.}
    \label{fig:S3}
\end{figure}

In the strong-$D$ phase, we use a unit cell consisting of six sublattices, as illustrated in Fig.~\ref{fig:1}(a). Using Eq.~\ref{eq:frame}, we obtain 
\begin{equation}
\begin{aligned}
    S_i^x S_j^y-S_i^yS_j^x& =-\sin(\Delta\phi) \,\tilde{S}_i^x \tilde{S}_j^x \\
    & -\cos(\theta_i)\cos(\theta_j)\sin(\Delta\phi) \,\tilde{S}_i^y \tilde{S}_j^{y} \\
    & -\sin(\theta_i)\sin(\theta_j)\sin(\Delta\phi)\, \tilde{S}_i^z \tilde{S}_j^z \\
    & \cos(\theta_j)\cos(\Delta\phi)\,\tilde{S}_i^x \tilde{S}_j^y  -\cos(\theta_i)\cos(\Delta\phi)\,\tilde{S}_i^y \tilde{S}_j^x\\
    & -\sin\theta_j \cos(\Delta\phi)\,\tilde{S}_i^x \tilde{S}_j^z 
 + \sin\theta_i \cos(\Delta\phi)\,\tilde{S}_i^z \tilde{S}_j^x \\
& +\cos\theta_i \sin\theta_j \sin(\Delta\phi)\,\tilde{S}_i^y \tilde{S}_j^z \\&
 +\sin\theta_i \cos\theta_j \sin(\Delta\phi)\,\tilde{S}_i^z \tilde{S}_j^y .
\end{aligned}
\end{equation}
and
\begin{equation}
\begin{aligned}
\vec{S}_i\cdot\vec{S}_j&=
\cos\Delta\phi\,\tilde S_i^x \tilde S_j^x \\
&+ \big(\sin\theta_i \sin\theta_j + \cos\theta_i \cos\theta_j \cos\Delta\phi\big)\,\tilde S_i^y \tilde S_j^y \\
 &+ \big(\sin\theta_i \sin\theta_j \cos\Delta\phi + \cos\theta_i \cos\theta_j\big)\,\tilde S_i^z \tilde S_j^z\\
&+ \cos\theta_j \sin\Delta\phi\,\tilde S_i^x \tilde S_j^y- \cos\theta_i \sin\Delta\phi\,\tilde S_i^y \tilde S_j^x
\\
&- \sin\theta_j \sin\Delta\phi\,\tilde S_i^x \tilde S_j^z + \sin\theta_i \sin\Delta\phi\,\tilde S_i^z \tilde S_j^x\\ &+ \big(\sin\theta_i \cos\theta_j - \sin\theta_j \cos\theta_i \cos\Delta\phi\big)\,\tilde S_i^y \tilde S_j^z \\
& + \big( \sin\theta_j \cos\theta_i-\sin\theta_i \cos\theta_j \cos\Delta\phi\big)\,\tilde S_i^z \tilde S_j^y,
\end{aligned}
\end{equation}
where $\Delta\phi=\phi_i-\phi_j$. After performing a HP transformation as in the main text, and Fourier transformation, we obtain the LSW Hamiltonian:
\begin{equation}
H^{\text{strong}}_m=\frac{1}{2}\sum_{\boldsymbol{k}} \begin{bmatrix}
\boldsymbol{\beta}_{\boldsymbol{k}}^{\dagger} \boldsymbol{\beta}_{-\boldsymbol{k}}
\end{bmatrix}
\left(\begin{array}{cc}
A_{\boldsymbol{k}}& B_{\boldsymbol{k}} \\
B_{-\boldsymbol{k}}^* & A^*_{-\boldsymbol{k}}
\end{array} \right)
\begin{bmatrix}
\boldsymbol{\beta}_{\boldsymbol{k}} \\
\boldsymbol{\beta}_{-\boldsymbol{k}}^{\dagger}
\end{bmatrix},
\end{equation}
where $\boldsymbol{\beta}_{\boldsymbol{k}}^{\dagger} \equiv 
\left[\beta_{1,\boldsymbol{k}}^{\dagger}, \ldots, \beta_{6,\boldsymbol{k}}^{\dagger}\right]$ 
denotes the bosonic operators for the $6$ sublattices, $A_{\boldsymbol{k}}$ and $B_{\boldsymbol{k}}$ for $\phi_2=\phi_1$ are defined by

\begin{widetext}
    \begin{equation}
    \begin{aligned}
        A_{\boldsymbol{k}}=&JS \left(\begin{array}{cccccc}
        -\sum_{i=1}^3z_{J,\boldsymbol{k}}(i) &g_{J,\boldsymbol{k}}(1,2)^*&0&g_{J,\boldsymbol{k}}(2,3)&0&g_{J,\boldsymbol{k}}(3,1)\\
    g_{J,\boldsymbol{k}}(1,2)&-\sum_{i=1}^3z_{J,\boldsymbol{k}}(i)&g_{J,\boldsymbol{k}}(3,3)^*&0&g_{J,\boldsymbol{k}}(2,1)^*&0\\
    0&g_{J,\boldsymbol{k}}(3,3)&-z_{J,\boldsymbol{k}}(1)-2z_{J,\boldsymbol{k}}(3)&g_{J,\boldsymbol{k}}(1,1)&0&g_{J,\boldsymbol{k}}(3,2)^*\\
    g_{J,\boldsymbol{k}}(2,3)^*&0&g_{J,\boldsymbol{k}}(1,1)^*&-z_{J,\boldsymbol{k}}(1)-2z_{J,\boldsymbol{k}}(2)&g_{J,\boldsymbol{k}}(2,2)&0\\
    0&g_{J,\boldsymbol{k}}(2,1)&0&g_{J,\boldsymbol{k}}(2,2)^*&-z_{J,\boldsymbol{k}}(1)-2z_{J,\boldsymbol{k}}(2)&g_{J,\boldsymbol{k}}(1,3)\\
    g_{J,\boldsymbol{k}}(3,1)^*&0&g_{J,\boldsymbol{k}}(3,2)&0&g_{J,\boldsymbol{k}}(1,3)^*&     -z_{J,\boldsymbol{k}}(1)-2z_{J,\boldsymbol{k}}(3)
        \end{array}\right)\\
        +&DS\left(\begin{array}{cccccc}
        -6z_{D,\boldsymbol{k}}&0&g_{D,\boldsymbol{k}}(1)^*&0&g_{D,\boldsymbol{k}}(1)&0\\
        0&-6z_{D,\boldsymbol{k}}&0&g_{D,\boldsymbol{k}}(2)^*&0&g_{D,\boldsymbol{k}}(2)\\
        g_{D,\boldsymbol{k}}(1)&0&-6z_{D,\boldsymbol{k}}&0&g_{D,\boldsymbol{k}}(1)^*&0\\
        0&g_{D,\boldsymbol{k}}(2)&0&-6z_{D,\boldsymbol{k}}&0&g_{D,\boldsymbol{k}}(2)^*\\
        g_{D,\boldsymbol{k}}(1)^*&0&g_{D,\boldsymbol{k}}(1)&0&-6z_{D,\boldsymbol{k}}&0\\
        0&g_{D,\boldsymbol{k}}(2)^*&0&g_{D,\boldsymbol{k}}(2)&0&-6z_{D,\boldsymbol{k}}\\
    \end{array}\right)+h\cos(\theta)\left(\begin{array}{cccccc}
    1&0&0&0&0&0\\
    0&1&0&0&0&0\\
    0&0&1&0&0&0\\
    0&0&0&1&0&0\\
    0&0&0&0&1&0\\
    0&0&0&0&0&1
       \end{array}\right), \\
    B_{\boldsymbol{k}}=&JS \left(\begin{array}{cccccc}
    0&l_{J,\boldsymbol{k}}(1,2)^*&0&l_{J,-\boldsymbol{k}}(2,3)^*&0&l_{J,-\boldsymbol{k}}(3,1)^*\\
    l_{J,-\boldsymbol{k}}(1,2)^*&0&l_{J,\boldsymbol{k}}(3,3)^*&0&l_{J,\boldsymbol{k}}(2,1)^*&0\\
    0&l_{J,-\boldsymbol{k}}(3,3)^*&0&l_{J,-\boldsymbol{k}}(1,1)^*&0&l_{J,\boldsymbol{k}}(3,2)^*\\
    l_{J,\boldsymbol{k}}(2,3)^*&0&l_{J,\boldsymbol{k}}(1,1)^*&0&l_{J,-\boldsymbol{k}}(2,2)^*&0\\
    0&l_{J,-\boldsymbol{k}}(2,1)^*&0&l_{J,\boldsymbol{k}}(2,2)^*&0&l_{J,-\boldsymbol{k}}(1,3)^*\\
    l_{J,\boldsymbol{k}}(3,1)^*&0&l_{J,-\boldsymbol{k}}(3,2)^*&0&l_{J,\boldsymbol{k}}(1,3)^*&0\end{array}\right)\\
    +&DS\left(\begin{array}{cccccc}
    0&0&l_{D,\boldsymbol{k}}^*&0&l_{D,-\boldsymbol{k}}^*&0\\
    0&0&0&l_{D,\boldsymbol{k}}&0&l_{D,-\boldsymbol{k}}\\
    l_{D,-\boldsymbol{k}}^*&0&0&0&l_{D,\boldsymbol{k}}^*&0\\
    0&l_{D,-\boldsymbol{k}}&0&0&0&l_{D,\boldsymbol{k}}\\
    l_{D,\boldsymbol{k}}^*&0&l_{D,-\boldsymbol{k}}^*&0&0&0\\
    0&l_{D,\boldsymbol{k}}&0&l_{D,-\boldsymbol{k}}&0&0
    \end{array}\right).
    \end{aligned}
    \end{equation}
\end{widetext}

The matrix elements for the Heisenberg interaction are given by
\begin{equation}
\begin{aligned}
g_{J,\boldsymbol{k}}(i,j) &= \left(x_{J,\boldsymbol{k}}(i)+y_{J,\boldsymbol{k}}(i)+\mathrm{i}2 p_{J,\boldsymbol{k}}(i) \right) f_{J,\boldsymbol{k}}(j)/2, \\
l_{J,\boldsymbol{k}}(i,j) &=(x_{J,\boldsymbol{k}}(i)-y_{J,\boldsymbol{k}}(i)) f_{J,\boldsymbol{k}}(j)/2, \qquad i,j=1,2,3,
\end{aligned}
\end{equation}
where \begin{equation}
\begin{aligned}
x_{J,\boldsymbol{k}}(i)&= \cos(\delta\phi_i)\\y_{J,\boldsymbol{k}}(i) &= \sin^2\theta + \cos^2\theta \cos(\delta\phi_i)\\p_{J,\boldsymbol{k}}(i)&= \cos\theta \sin(\delta\phi_i)\\z_{J,\boldsymbol{k}}(i)&= \sin^2\theta \cos(\delta\phi_i) + \cos^2\theta\\
f_{J,\boldsymbol{k}}(1) &= e^{-i\left(\tfrac{k_1}{3}-\tfrac{k_2}{3}\right)}\\f_{J,\boldsymbol{k}}(2) &= e^{-i k_1/3}\\f_{J,\boldsymbol{k}}(3) &= e^{-i k_2/3}
\end{aligned}
\end{equation}

and

\begin{equation}
    \delta\phi(i) =
\begin{cases}
0, & i=1, \\
-\tfrac{2\pi}{3}\,\mathrm{sgn}(D), & i=2, \\
+\tfrac{2\pi}{3}\,\mathrm{sgn}(D), & i=3 .
\end{cases}
\end{equation}

The matrix elements for the DMI are given by
\begin{equation}
\begin{aligned}
g_{D,\boldsymbol{k}}(1)&=(x_{D,\boldsymbol{k}}+y_{D,\boldsymbol{k}}+\mathrm{i}2p_{D,\boldsymbol{k}})f_{D,\boldsymbol{k}}/2\\
g_{D,\boldsymbol{k}}(2)&=(x_{D,\boldsymbol{k}}+y_{D,\boldsymbol{k}}+\mathrm{i}2p_{D,\boldsymbol{k}})f_{D,\boldsymbol{k}}^*/2\\
l_{D,\boldsymbol{k}}&=(x_{D,\boldsymbol{k}}-y_{D,\boldsymbol{k}})f_{D,\boldsymbol{k}}/2,
\end{aligned}
\end{equation}
where \begin{equation}
\begin{aligned}x_{D,\boldsymbol{k}}&=-\sin(\mathrm{sgn}(D)2\pi/3)\\y_{D,\boldsymbol{k}}&=-(\cos(\theta))^2\sin(\mathrm{sgn}(D) 2\pi/3)\\p_{D,\boldsymbol{k}}&=\cos(\theta)\cos(\mathrm{sgn}(D)2\pi/3)\\z_{D,\boldsymbol{k}}&=-(\sin(\theta))^2\sin(\mathrm{sgn}(D)2\pi/3)\\
f_{D,\boldsymbol{k}} &= e^{i\left(\tfrac{2k_1}{3}-\tfrac{k_2}{3}\right)}+e^{i\left(-\tfrac{k_1}{3}+\tfrac{2k_2}{3}\right)}+e^{i\left(-\tfrac{k_1}{3}-\tfrac{k_2}{3}\right)}.\end{aligned}
\end{equation}
When $h=0$, one has $\cos\theta=0$, which enforces $p_{D,\boldsymbol{k}}=p_{J,\boldsymbol{k}}=0$ and thus $\boldsymbol{H}_{\boldsymbol{k}}=\boldsymbol{H}_{-\boldsymbol{k}}^{*}$. For $h\neq 0$, however, $\cos\theta\neq 0$ so that $p_{D,\boldsymbol{k}}$ and $p_{J,\boldsymbol{k}}$ become nonzero, leading to $\boldsymbol{H}_{\boldsymbol{k}}\neq\boldsymbol{H}_{-\boldsymbol{k}}^*$ and enabling nonzero Chern numbers.

Diagonalizating the Hamiltonian yields six magnon bands, as shown in Fig.~\ref{fig:S2}. The magnon bands along high symmetric line are plotted in Fig.~\ref{fig:2}. Figure.~\ref{fig:S3} shows the band gaps (minimum of energy differences), which characterize the topological phase transitions. We determine the parameter values $D$ and $h$ at which band touching occurs by computing minimum energy differences, and plot them by the colorful curves in Fig.~\ref{fig:3}. The band gap between the upper three and the lower three bands, namely $E_4-E_3$ increases with increasing $h$ but decreases with increasing $D$, as shown in Fig.~\ref{fig:S3}(c). This behavior leads to a decrease of $\kappa_{xy}$ in Fig.~\ref{fig:4}(b) and an increase in Fig.~\ref{fig:4}(c).

\end{document}